\documentclass[12pt]{article}
\usepackage{graphicx}

\begin{document}

 \centerline{\large\bf Mixed scenario of the charged}
 \centerline{\large\bf helium surface reconstruction}

\vskip 3 mm \centerline{V.Shikin, E.Klinovaya}

\centerline{\sl\small Institute of Solid State Physics of RAS,
Chernogolovka, Moscow distr., 142432, Russia}

\vskip 3 mm

\begin{abstract}
Discussed in the paper is the mixed scenario of charged liquid
surface reconstruction when the surface 2D charge density is close
to the saturation. The basic building block of arising honeycomb
structure is shown to be a modified multielectron dimple.
\end{abstract}


measurements and simulations;\\ 68.03.Kn Dynamics (capillary
waves) \vskip 3mm

One of the instabilities studied in classical hydrodynamics is the
the Frenkel-Tonks (FT) instability \cite{1, 2} occurring in a
threshold manner at the charged liquid surface and resulting in
its deformation. A distinctive feature of the FT instability
compared with other decay processes: (Rayleigh instability of the
cylindrical jet \cite{3}, Karman trace behind the moving cylinder
(sphere) \cite{4}, instability of the interface between two liquid
media moving relative to each other \cite{5}, Taylor vortex
instability of viscous liquids between two rotating coaxial
cylinders \cite{5}, etc. is the possibility of stopping the decay
process after which a new metastable state with a finite
corrugation amplitude is formed (i.e., a reconstruction occurs).
At different times this phenomenon was considered by many authors
for dielectric and charged interfaces \cite{6} - \cite{28}.
However, many important details of this process have only benn
understood quite recently. In the FT instability, only the initial
part of the problem (calculation of the dispersion law and it
stability analysis) has a universal nature. Further analysis is
requires different treatments for the dielectric and charged
boundaries. In contrast to the dielectric interfaces, the latter
case allows formation of the solitons of a special kind, the
multielctron dimples, \cite{21,22}. A less obvious, but actually
more important factor resulting in different behaviour of nuetral
and charged surfaces is the violation of homogeneity of boundary
conditions (the surface charge distribution along the interface
can become discontinuous). Earlier Mel'nikov and Meshkov showed
\cite{23,25} that the boundary condition inhomogeneity inevitably
occurs in the problem of the charged helium surface
reconstruction. Therefore, its analysis should account for arising
inhomogeneities.

The first of all inhomogeneous reconstruction scenarios we are
aware of substantially employing the elements of equipotential
reconstruction theory \cite{11, 12, 13, 15, 20}, and hence the
important for this formalism low electron population $\nu\ll 1$ of
the liquid surface was proposed by authors of Ref. \cite{23,25}.
Omitting the details, we only note that ``bald'' areas arise in
their treatment as a triangular lattice (whose primitive vectors
are equal to the capillary lengths $a$) of small spots having the
shape of circles of radius $ R^*\ll a, \ a=\kappa^{-1}$, growing
from zero as the supercriticality rises. Unfortunately, the theory
given in Refs. \cite{23,25} does not contain individual solitons
(which are typical of the charged helium surface reconstruction
observed for $\nu\ll 1$) among its solutions. In addition, the
maximum surface curvature in the central parts of nuetral spots
has the sign opposite to that observed experimentally \cite{27}.
These circumstances hamper evaluation of the feasibility of the
model proposed in Ref.  \cite{23,25}. Later, it was found that in
the range of $\nu\ll 1$ the details of the reconstruction are much
more naturally described in terms of individual multielectron
dimples \cite{28}. Here and below the boundary population with
mobile charges $\nu$ is defined as the ratio
\begin{equation}
 \nu=n_s/n_s^{max},\quad 2\pi( e n_s^{max})^2=\sqrt{\alpha \rho g},
\label{1}
\end{equation}
where $\rho$ and $\alpha$ are the density and surface tension, $g$
is the acceleration due to gravity, $e$ is the electron charge,
$\kappa=\sqrt{\rho g/\alpha}$ is the liquid capillary constant.

In the opposite, most suitable for observations limit $\nu\le 1$
where the equipotential language \cite{15,20} and its modification
\cite{23,25} do not work at all, it was suggested to model the
corrugation with inhomogeneous electrostatic potential
distribution by a periodic system of multielectron dimples
\cite{26}. The analysis of Ref. \cite{26} is based on tight
binding approximation. The structure is assumed to consist of
roughly estimated free multielectron dimples. By further assuming
these quasiparticles to be point-like, the authors model the
periodic lattice of charged dimples by employing the results
derived for classical Coulomb crystal consisting of point charges
\cite{29}.  The dimple crystal spacing is approximately equal to
the capillary length while its advantage compared with the
homogeneous 2D charge distribution is treated in terms of
correlations \cite{29}. To comment on the results of Ref.
\cite{26}, we note that the single dimple problem accurately
solved in Ref. \cite{24} yields for the conditions of Ref.
\cite{26} the charged spot radius considerably exceeding the
lattice spacing given in Ref. \cite{26} (according to Ref.
\cite{24}, here $R/a\simeq 1.5$). The critical conditions for its
development is higher than the dynamic stability threshold.
Finally, we believe the correlation effects \cite{29} are not
directly related to the Coulomb part of the reconstruction
problem. Thus, the dimple picture of a periodic reconstruction,
being almost the only realistic in the range of $\nu\le 1$,
contains in the formulation given in Ref. \cite{26} a number of
disputable concepts whose presence should be at least understood
and which should be modified. The details of such an analysis are
given in the present report.

1. From our point of view, most questionable are suggestions of
authors of Ref. \cite{26} on the Coulomb part of the problem. It
is assumed that the development of a periodic corrugation is
strongly assisted by the well-known gain in energy occurring when
the 2D charged systems goes from the gaseous (liquid) state to the
crystalline one (Wigner crystal). Such a gain of the correlation
origin indeed takes place (e.g. see Ref. \cite{29}) as revealed by
analysis of the difference between the average Coulomb energy of
(i) likely charged point-like particles $<V_c^l(r)>$ uniformly
distributed along the surface with the average density $n_s$, and
(ii) those arranged in a lattice $<V_c^s(r)>$ with the same
average density
\begin{equation}
\delta<V_c(r)>=<V_c^l(r)>-<V_c^s(r)>\simeq +c_s e^2\sqrt{n_s}.
\label{2}
\end{equation}
Here the constant $c_s$ of the order of unity depends on the
lattice type. By employing a formal analogy between the structure
of a classical Wigner solid with one electron per cell and the
corrugation having spacing $b$ and the charge $q_0$ per cell, it
is possible to assign to the Coulomb part of the reconstruction
problem (as it was done in Ref. \cite{26}) the energy gain of the
order of
\begin{equation}
 \delta<W_c(r)>\simeq +c_s q_0^2/b >0
\label{3}
\end{equation}

However, the hypothesis (\ref{3}) can actually be checked and, at
least for the employed model of a corrugation involving formation
of a set of one-dimensional conducting threads in the practically
interest case $\nu = 1$ is not confirmed. Detailed calculations
(which can be found in the Appendix) lead to the energy whose sign
is opposite to that in equation (\ref{3}):
\begin{equation}
  \delta<W_c(r)>=\frac{Q^2}{2}\left(1/\bar C-1/\tilde C \right)< 0
\label{4}
\end{equation}
\begin{equation}
 \bar C=\frac{S}{4\pi d}, \quad \tilde C=\frac{S}{4\pi
d+4 a\ln{(a/\pi R)}}\label{5}
\end{equation}
where $S$ is the capacitor area, $R$ is the single thread radius,
$2a$ is the corrugation period. In addition to the sign which is
opposite to that of Eq. \ref{3}), the formula (\ref{4}) has one
more qualitatively important property: here the energy gain
depends explicitly on the length $R$ measuring the charges
localization length in a separate groove. This length is
determined self-consistently by requiring the minimum of the
corrugated system total energy. In the approximation used in Ref.
(\ref{3}), the Coulomb part of the problem which does not contain
$R$ slips out of the self-consistency condition which should not
be allowed. In the rest of the paper we follow the definition of
$\delta<W_c(r)>$ given by Eq. (\ref{4}).

\begin{figure}
\centering
\includegraphics[width=99 mm]{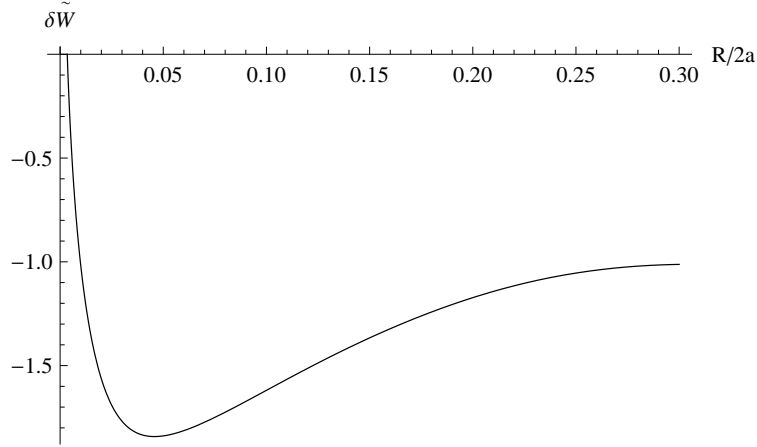}
\caption{\label{pic1}Energy gain $\delta\tilde W$ (\ref{7}) as a
function of the electron spot radius. The negative minimum
corresponds to the optimal value of $R/a$.}
\end{figure}

\begin{figure}
\centering
\includegraphics[width=99 mm]{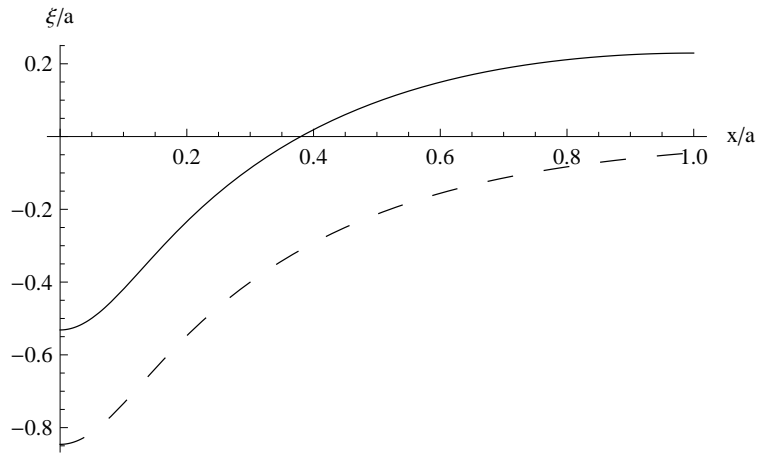}
\caption{\label{pic2} Fig. 2. One-dimensional dimple profiles for
a single free dimple (dashes) and a periodic set of dimples (solid
line).}
\end{figure}

2. Here are some remarks before the study of the corrugation
problem for $\nu\le 1$. First of all, the total charge $Q$ at the
liquid-vapor interface with area $S \sim L^2$ is constant:
\begin{equation}
    Q=const= e n_s^{max} L^2 \label{6}
\end{equation}

Perhaps a trivial point, but the problem formulation should for
definiteness contain some comments on the initial stage of the
decay. It develops within the dynamic equipotential scenario
\cite{15}, until under the conditions of Eq. (\ref{6}) the growth
of the corrugation amplitude results in the break of continuity in
the charge distribution. In the limit of small $\nu$ the
inevitable appearance of continuity breaks was noted by authors of
Refs. \cite{23,25}. However, their reasoning is essentially
universal in nature and is applicable for arbitrary values of
$\nu$. If one assumes that the spatial period of developing
corrugation is not changed with time (in the general theory of
spinodal kinetics relevant to the problem considered here this
assumption is not true \cite{30}), the net charge splits into a
system of $N$ individual clusters of charge $Q_0$ per unit length
\begin{equation}
N=L/2a, \quad   Q_0\simeq 2 a e n_s^{max} \quad  \label{7}
\end{equation}
each of which is the nucleus of a separate one-dimensional (for
simplicity) dimple residing at its center. It should be noted that
the unstable state could decay in a purely fluctuational (binodal)
way just as in the case of $\nu\ll 1$ (see Ref. \cite{28}).
However, in the general case of arbitrary $\nu$ the dimple charge
remains undefined in this scenario. The decay having a spinodal
initial and binodal final stages (we call it ``intermediate'')
does not contain this uncertainty.

In the problem of finding the corrugated surface state in the
critical electric field $E_+^{max}$ with the charge $Q_0$
(\ref{3}) per cell, one should verify the ability of the deformed
helium surface to hold the critical charge (perhaps tking into
account the non-linear phenomena) and check the validity of
inequality
\begin{equation}
\bar W(E_+^{max},Q)\ge \tilde W(E_+^{max},Q_0,b),\quad
E_+^{max}=4\pi e n_s^{max} \label{8}
\end{equation}
where $n_s^{max}$ is taken from Eq. (\ref{1}), $\bar
W(E_+^{max},Q)$, $\tilde W(E_+^{max},Q_0,b)$ are the total
energies of the uniform and corrugated helium surface states, $b$
is the lattice spacing which is generally different from the
capillary length $a$. There are no reasons to require coincidence
of energies in Eq. (\ref{8}) since we do not consider the actual
binodal transition from a flat to corrugated state. Nevertheless,
it is desirable to have the energy $\tilde W(E_+^{max},Q_0,b)$ to
be less than $\bar W(E_+^{max},Q)$, since otherwise the geberal
picture of the system evolution could not be reasonably
interpreted.

Bearing in mind the above arguments and results derived in the
Appendix, the problem of one-dimensional corrugation is reduced to
calculating the energy difference $\bar W(E_+^{max},Q)$~- ~$\tilde
W(E_+^{max},Q_0,b)$ (\ref{8}). Note that the one-dimensional
scenario is not only a convenient model but also an experimentally
observed state of the corrugated surface \cite{27}. In addition,
this is the least stable type of corrugation. Laplace pressure
stabilizing the picture is twice as low in the one-dimensional
case as in two dimensions. The existence of a solution yields an
estimate from below for the corrugation stability.

The initial energy functional for a one-dimensional corrugation
(per unit thread length)
\begin{equation}
 \delta\tilde W(E_+^{max},R,a)= \int\limits_{-a}^{+a} d x \left
[\frac{\alpha}{2}[(\nabla\xi)^{2} + \kappa^{2} \xi^{2}]+
eE_+^{max}\delta n(x)\xi(x)  \right ]+\delta<W_c(r)>, \label{9}
\end{equation}
$$
\delta<W_c(r)>= 2e^2n_s^2 a^2\ln{\frac{a}{\pi R}}
$$
$$
\xi(x)=\zeta(x)-<\zeta>, \quad \delta n(x)=n(x)-n_s
$$
represents the system energy counted from the energy of uniformly
charged flat liquid surface and written for one lattice cell. Its
Coulomb part is taken in the form of Eq. (4) allowing variational
solution for the problem of finding the length $R$ if the charge
density $n(x)$ in a periodic system of dimples is split into
individual stripes each having some density distribution $n_0(x)$
(for example, a Gaussian one)
\begin{equation}
 n(x)=\sum_l n_0(x-2la),\quad n_0(x)=n_0 \exp{(-x^2/R^2)},\quad
\sqrt{\pi} n_0=\frac{n_s a}{erf(a/R) R }\label{10}
\end{equation}
($erf(x)$ is the known special function) normalized according to
Eq. (\ref{7}). Average values of the variables $\zeta(x)$, $n(x)$
are subtracted from their initial definitions according to the
general structure of the functional (\ref{9}).

The functional (\ref{9}) has the standard for the theory of
multielectron dimples structure. The deformation-related part
written in square brackets makes the reduction of the parameter
$R/a$ as well as the growth of liquid surface deformation due to
local electron pressure energetically favorable. The Coulomb
component $\delta<W_c(r)>$ hinders this process. The interplay
between these factors results in the appearance of a negative
minimum in the $\delta\tilde W(E_+^{max},R/a)$ (\ref{9})
dependence which determines the equilibrium value of the parameter
$R/a$. To get the numerical values, one should solve the liquid
surface mechanical equilibrium equation for $\xi(x)$ assuming
$n(x)$ (\ref{10}) to be a known function, then calculate the
integrals (\ref{9}) with these distribution, and finally obtain
the energy (\ref{9}) dependence on the variational parameter
$R/a$. For the external parameters corresponding to $\nu=1$ this
picture is presented in Fig. \ref{pic1}.

The functional minimum is reached at $R/a=0.1$ and is negative. In
other words, the corrugated state is more favorable than the flat
one. The corrugation is stabilized in the range of $R/a \ll 1$,
although the obtained numbers should only be considered as a rough
estimate because the values of $\nabla \xi(x)\le 1$ are not
parametrically small, as follows from the data on the
one-dimensional dimple profile presented in Fig. \ref{pic2}). Also
shown for comparison in the same figure is the individual free
dimple profile. Obviously, the linear theory corresponding to the
bilinear energy functional (\ref{9}) requires quantitative
corrections (it follows from the figure that the inequality
$\nabla \xi(x)< 1$ used to derive Eq. (\ref{9}) is only poorly
met). However, qualitatively the mixed reconstruction scenario is
self-consistent (starting from the assumption on the presence of
continuity breaks in the equilibrium distribution $n(x)$ we
finally obtain the picture confirming the initial assumptions of
the theory).

To sum up, we conclude that the mixed scenario of the charged
liquid surface reconstruction under the conditions of $\nu=1$
involving the formation of a periodic system of charged dimples
has a self-consistent nature.  The period of arising structure is
determined by the dynamical instability of charged liquid surface.
Development of this instability divides the surface charges into a
periodic system of clusters localized in every  cell of a
one-dimensional or a honey-comb corrugation. When approaching the
metastable state, the charged cell cores becomes rather small
($R/a \ll 1$) while the corrugation amplitudes is of the order of
the capillary length (stiff reconstruction regime according to the
classification of Ref. \cite{11}). The mixed scenario reveals no
hysteresis phenomena typical of nonlinear equipotential theories
(which is not surprising because of the employed linear
approximation (9) for energy).

Formally, the Coulomb energy should be treated in a consistent
way. The field-theoretical approach allows one to avoid paradoxes
at both quantitative and qualitative levels. In particular, in the
considered problem there are no serious reasons for manifestation
of correlation phenomena governing transition to crystalline
ordering of the Wigner type.

This work was partly supported by the RFBR grant 09-02-00894а.

 \centerline{\sl\small Appendix}

\renewcommand{\theequation}{П\arabic{equation}}
\setcounter{equation}{0}

The statements from Ref. \cite{28} concerning the properties of
the difference
\begin{equation}
 \delta<V_c(r)>= <V_c^l(r)>-<V_c^s(r)>
\end{equation}

are based on the analysis of the expression
\begin{eqnarray}
 \delta<V_c(r)>&=&e^2\lim_{\vec x\to 0}\sum_{l}\left[\frac{1}{|\vec x-\vec
x(l)|}-\frac{1}{|\vec x|}\right]\label{P1}\\
\vec x(l)&=&l_1\vec a_1+l_2\vec a_2,
\end{eqnarray}
where $\vec a_1$ и $\vec a_2$ are the primitive translation
vectors in the considered lattice.

The sums in Eq. (\ref{P1}) diverge already because the classical
proper energy of every electron in the definition (\ref{P1}) is
infinite. However, the Fourier analysis and the important
condition of the system neutrality as a whole (which is insured by
introducing the screening electrode) allow to reduce the
difference of the sums to a finite expression of the form (e.g.,
see Ref. \cite{29})
\begin{equation}
 \delta<V_c(r)>=N c_s e^2\sqrt{a_c}/2, \label{P2}
\end{equation}
where $N$ is the total numcer of elctrons in the 2D system,
$a_c\simeq n_s^{-1}$ is the unit cell area, $c_s$ is a constant of
the order of unity depending on the lattice type.

An elegant and rather general result (\ref{P2}) gives an idea on
the nature of Coulomb crystallization of electrons above helium
yielding the crystal ground state energy in this problem compared
to the gaseous (liquid) phase.

Let us now turn to the known electrostatic problem \cite{31} on
the field and energy of a set of charged threads above a
conducting screen, i.e. the problem of a flat capacitor with a
solid or mesh-like electrodes. Here the electrostatic potential
has the form
\begin{equation}
\varphi(z)=-2a\sigma \ln  \left( \frac{\sin{[\pi(z-id)/a]}}
{\sin{[\pi(z+id)/a]}} \right),\quad z=x+iy  \label{P3}
\end{equation}
where $a$ is the lattice spacing, $\sigma$ is the mesh charge per
unit area, $d$ is the distance between the mesh and the screen.
Just as in (\ref{P1}), the divergencies at infinity cancel each
other. The filed is concentrated within the capacitor, the mesh
filed becoming uniform at distances of the order of its period
$2a\ll2d$ where $2d$ is the distance between the plates. The total
electrostatic energy (the counterpart of Eq. (\ref{P1}))
calculated in the field-theoretical approach (squared electric
field integrated over the capacitor volume) proves to be divergent
due to the potential  $\varphi(z)$ (\ref{P3}) singularity at the
threads. This property requiring cutoff at the thread radius $R\ll
a$ does not appear in the final result (\ref{P2}).

Capacities $\bar C$, $\tilde C$ per unit system area after using
Eq. (\ref{P3}) can be calculated to be
$$
\bar C=\frac{1}{4\pi b}, \quad \tilde C=\frac{1}{4\pi
b+2a\ln{(a/2\pi R)}}.
$$
Here $R\ll a/2\pi$ is the radius of a single thread.

The counterpart of expression (\ref{P1}) is obvious:
\begin{equation}
\delta<W_c(r)>=\frac{Q^2}{2}(1/\bar C-1/\tilde C) \label{P5}
\end{equation}

This difference has the sign opposite to that of Eq. (\ref{P2})
and depends on particular characteristics of a single thread.
Omitting any comments on possible reasons of the qualitative
difference between Eqs. (\ref{P2}) and (\ref{P5}) we shall employ
in the main body of the paper the Coulomb part defined by Eq.
(\ref{P5}) since it contains information on the internal Coulomb
energy of every charged thread which is absent from Eq.
(\ref{P2}).

\end{document}